# Accelerating Neutron Scattering Data Collection and Experiments Using AI Deep Super-Resolution Learning

Ming-Ching Chang[1], Yi Wei[1], Wei-Ren Chen[2], Changwoo Do[2†]

[1]University at Albany, State University of New York, NY, USA.

[2]Neutron Scattering Division, Oak Ridge National Laboratory, Oak Ridge, USA.

Pursuing hardware upgrades to provide brighter beams for material studies has been the paramount goal of every neutron scattering facility. Here, we present an alternative route to circumvent the limitation of neutron flux using recent advancements in artificial intelligence (AI), namely image super-resolution deep convolutional neural network (CNN). The feasibility of accelerating data collection has been demonstrated by using small angle neutron scattering data collected from the Extended Q-range Small Angle Neutron Scattering (EQ-SANS) instrument at Spallation Neutron Source (SNS). Data collection time can be reduced by increasing the size of binning of the detector pixels at the sacrifice of resolution. High-resolution scattering data is then reconstructed by using AI deep super-resolution learning method. This technique can not only improve the productivity of neutron scattering instruments by speeding up the experimental workflow but also enable capturing kinetic changes and transient phenomenon of materials that are currently inaccessible by existing neutron scattering techniques.

Keywords: neutron scattering, super-resolution, deep neural network, small angle scattering, neutron facility

# 1. Introduction

It is now a common practice of using neutron scattering techniques to investigate atomic-scale structure and dynamics. They possess unique advantages over other scattering techniques such as x-ray or light, with the exceptional penetration power, free from radiation damage, and the ability to selectively highlight specific parts of materials via isotope labelling(Lindner and Zemb, 2002; Richter et al., 2005). However, currently available neutron flux is significantly lower than that of photons generated by modern synchrotron radiation or free-electron laser sources. To make things more challenging, strong penetration power comes with much smaller cross-section for neutrons. Therefore, lengthy measurement time, usually tens of minutes to tens of hours even at facilities with most powerful neutron sources, is required to obtain high-quality data with satisfactory signal-to-noise ratio.

The requirement of long measurement time not only limits the productivity of the neutron user facility but also has made real-time characterization experiments very challenging for neutron scattering techniques. For example, understanding dynamic response under external stimuli has been an active research field of non-equilibrium materials(Milne et al., 2014; Narayanan et al., 2017). To understand the structural and dynamical changes *in-situ* or *operando*, as in materials undergoing phase transition, external stimuli, or mechanical deformation, a snapshot of a material's structure and dynamics need to be taken periodically on a time scale relevant to the dynamic changes. While time-synchronised sample environment in combination with event-based processing of neutron data has enabled observation of fast dynamic changes even with neutrons(Granroth et al., 2018), irreversible phenomena can still be measured only with high enough flux. Therefore, minutes to tens of seconds of time resolution has been the best time resolution which can provide good signals for studying irreversible process using neutron scatterings(Bruetzel et al., 2018; Lund et al., 2007; Sauter et al., 2015). This is at least one or two orders of magnitude slower time resolution that x-ray scattering can achieve(Bruetzel et al., 2018; Sauter et al., 2015; Vegso et al., 2017). Considering the unique capabilities that neutrons have, enabling fast data collection using neutrons will bring new science opportunities in materials research. Hardware investment (constructing a bigger accelerator and/or using a brighter source) to solve these aforementioned problems is impractical, as the return on investment is not cost effective. The spallation source has begun to reach its practically attainable flux limit and there are still various challenges left for the new type of neutron production such as an inertial fusion method becomes realized(Taylor et al., 2007). In addition, neutron flux will still remain insufficient to measure weakly scattering samples or to capture faster dynamic processes requiring shorter counting time.

One way to speed up the measurement while keeping good counting statistics per detector pixel is to reduce the number of detector pixels and increase individual pixel area. More neutrons per pixel will be collected if the area of the individual pixel is increased. Therefore, the same level of signal to noise ratio per individual pixel can be achieved much faster than high-resolution data collection which uses smaller individual pixel area and more pixels. In short, measurement speed is gained at the sacrifice of resolution. However, if high-resolution detector images can be recovered from the given low-resolution input, this will result in effective speed-up of the neutron experiments without compromising resolution. Super-resolution is an active research topic in digital image processing and computer vision(Wang et al., 2019), where mostly nature (electro-optical) images are studied. Here we adopt the state-of-the-art super-resolution algorithm to neutron scattering data. The aim is to recover a high-resolution image from its low-resolution signal. Due to physical acquisition limitations in image formation, high-resolution images are in general harder (or require longer time) to capture. Thus, super-resolution technique can be applied to improve the efficiency in the acquisition of high-resolution information from merely the observed low-resolution signals. Super-resolution is an ill-posed problem, since a single pixel in low resolution image could map to multiple pixels in its high-resolution counterparts. A key assumption of many super-resolution techniques is that the high frequency pattern is redundant and can be easily reconstructed from low frequency components. Existing super-resolution methods can be organized into three categories: (i) edge-based methods, (ii) image statistical methods, and (iii) example patch-based methods. Refer to(Wang et al., 2019) for a thorough review. Example patch-based methods have achieved good performance in traditional methods(Dong et al., 2011; Kwang In Kim and Younghee Kwon, 2010; Yang, 2010; Yang et al., 2013). Sparse dictionary learning (or sparse coding) is widely used in these methods, which assumes that the image signal can be represented by a dictionary of representation atoms. This way, the redundant patterns of high-frequency signals can be represented by dictionary atoms, and the correspondence between low-resolution and high-resolution representations is learnt through the sparsity-based formulation(Dong et al., 2011; Yang, 2010).

Recently, due to the booming popularity of Deep Learning and in Artificial Intelligence(LeCun et al., 2015), Deep Neural Networks (DNN) based methods can achieve excellent performance in many research fields including image analysis, healthcare, and natural language processing. Specific to our problem here, deep Convolutional Neural Network (CNN) based image super-resolution methods have shown promising results in enhancing low-resolution or noisy images by learning from sufficiently large training examples(Krizhevsky et al., 2012; Shi et al., 2016a). DNN models are known to be capable for learning superior features that can improve the reconstruction accuracy in super-resolution. Several deep neural networks based super-resolution methods(Chen and Pock, 2017; Dong et al., 2016; Shi et al., 2016a) with end-to-end learning have shown superior performance than the traditional sparse coding methods based on hand-crafted features. The basic idea of these methods is to learn a good feature representation in the top several layers of neural networks and map the low-level feature to high-resolution signal space. The whole process is trained end-to-end with little pre/post-processing beyond the optimization. We have recognized a great potential of applying the deep learning image super-resolution approach to accelerate neutron scattering data collection, particularly small angle neutron scattering data to begin with, by taking advantage of large data sets available at EQ-SANS (SNS, Oak Ridge).

In this work, we apply the one of the well-known super-resolution convolutional neutral networks(Shi et al., 2016a) to predict high-resolution scattering images from the low-resolution scattering inputs. By grouping detector pixels, good statistics data can be obtained at much faster rate. Therefore, successful super-resolution method can effectively accelerate the experimental time. A reasonably trained super-resolution neutral network is demonstrated by using randomly selected large neutron data sets under identical instrument configuration from EQ-SANS at SNS.

## 2. Methods

### 2.1. Data preparation

We prepare the SANS data collected from EQ-SANS at SNS to train a deep image super-resolution neutral network. As of August 2018, EQ-SANS has collected more than 80,000 measurements including both transmission and scattering data at various configurations. Each instrument configuration such as sample to detector distance, choices of beam slits, sample apertures, and wavelength bands introduce unique resolution function and $q$-ranges, where the wavevector $q$ is defined as $q = 4\pi \sin(\theta)/\lambda$ with scattering angle $\theta$ and wavelength $\lambda$. These are critical parameters that determines the broadening of scattering peaks and typically observed scattering patterns. For reasonable trainings, pairs of scattering images both at low-resolution and high-resolution are required. In order to isolate the effect of noise in the data, scattering images that have more than 5 million total neutron counts are preselected. Data sets are further filtered by the most widely used instrument configuration, which is 4m sample-to-detector distance and a wavelength band with 2.5A as a minimum wavelength. As a result, total 5,573 scattering data have been selected and reduced in two resolutions (120x120 and 30x30) following standard SANS data reduction procedure using MantidPlot described elsewhere(Heller et al., 2018). At this stage, the time of flight data have been reduced to static 2D data. During the process of the data, material and science-specific data were all removed from the metadata. Resulting three column data (intensity, $q_x$, $q_y$) have been converted into 2-D array, where the value of each element is the scattering intensity at the position of $q_x$ and $q_y$. We filled the element with 0 if the value of that position is missing.

## 2.2. Pre-processing

The intensity of input data is of large range (0 < intensity < ~ 51,000), where most of the intensity value are greater than 10,000. Data points of low intensity (less than 100) are sparse. Unbalanced data value can cause backpropagation optimization hard to converge during the training of neural networks. It can also reduce the accuracy of network outputs. Therefore, we applied log normalization method on data to rescale it to a smaller range.

$$x' = \ln(x + \varepsilon),$$

where $\varepsilon$ is set to make the base positive, and we use $\varepsilon = 3$ in our experiment. After rescaling, the range of input is (0, 16). After super-resolution, the output of neural network will be rescaled to its original range by an inverse operation to obtain a value in its original range.

## 2.3. Neural Network Model

In this work, we adopt the Efficient Sub-Pixel Convolutional Neural Network (ESPCN)(Shi et al., 2016a) for our experiment due to its good performance and time-efficiency. The ESPCN takes a low-resolution (LR) image (signal) $\mathbf{I}^{LR}$ with tensor size $H \times W \times C$ as input where $C$ is the colour channel and try to estimate the high-resolution (HR) ground truth $\mathbf{I}^{HR}$ with a specified upscaling ratio $r$. We denote the output of the neural network as $\mathbf{I}^{SR}$, which has the same tensor size with $\mathbf{I}^{HR}$ as $rH \times rW \times C$. The architecture of ESPCN is shown in **Fig. 1**, it consists of two parts: the first $L-1$ convolutional layers are applied to $\mathbf{I}^{LR}$ to learn the feature representation of input signal, then a sub-pixel convolution layer upscales the low-resolution feature maps to produce $\mathbf{I}^{SR}$.

The first $L-1$ convolutional layers can be represented as follows:

$$f^1(\mathbf{I}^{LR}; W_1, b_1) = \phi(W_1 * \mathbf{I}^{LR} + b_1)$$

$$f^l(\mathbf{I}^{LR}; W_{1:l}, b_{1:l}) = \phi(W_l * f^{l-1}(\mathbf{I}^{LR}) + b_l),$$

where $W_l, b_l, l \in [1, L-1]$ are learnable network weights and bias for each convolutional layer. $\phi(\cdot)$ is the nonlinear activation function. The Rectified Linear Unit (ReLU) used in the network as activation function is:

$$\text{ReLU}(x) = \max(0, x).$$

After the first $L-1$ convolutional layers, $\mathbf{I}^{LR}$ is represented by a $H \times W \times r^2C$ tensor. To generate the desired high-resolution image $\mathbf{I}^{SR}$ from LR feature maps, a sub-pixel convolution layer is used of a deconvolutional layer as in the reference(Shelhamer et al., 2017). Both proposed sub-pixel convolution layer and deconvolutional layer can map the same feature maps in LR space to HR space, however the sub-pixel convolution layer contains more parameters in convolution filters, thus with more representation capability in upsampling the input signal with the same computational speed. The last sub-pixel convolution layer is formulated as:

$$\mathbf{I}^{SR} = f^L(\mathbf{I}^{LR}) = PS(W_L * f^{L-1}(\mathbf{I}^{LR}) + b_L),$$

where PS is a periodic shuffling operator that rearranges the elements of a $H \times W \times r^2 C$ tensor to a tensor with shape $rH \times rW \times C$ which is the estimated super-resolution image $\mathbf{I}^{SR}$. Detailed implementation can be found in references(Shi et al., 2016a, 2016b). To train the parameters of network, a pixel-wise mean squared error (MSE) loss is applied as the objective function to measure the reconstruction errors:

$$\ell(W_{1:L}, b_{1:L}) = \frac{1}{r^2 HW} \sum_{i=1}^{rH} \sum_{j=1}^{rW} \left( \mathbf{I}_{i,j}^{HR} - f_{i,j}^L(\mathbf{I}^{LR}) \right)^2.$$

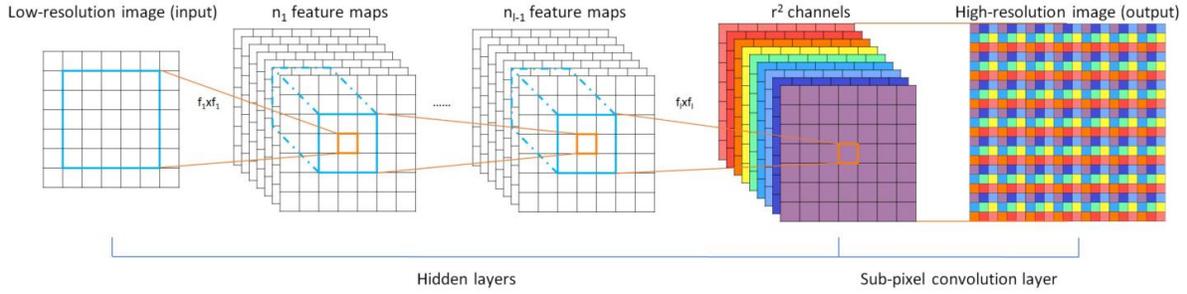

**Fig. 1**. Architecture of the efficient sub-pixel convolutional neural network (ESPCN)(Shi et al., 2016a). Figure adapted from the reference(Shi et al., 2016a). The top two convolution layers extracts feature maps, while the last sub-pixel convolution layer aggregates the feature maps from low-resolution space to build the super-resolution image.

**2.4. Training setting**

For the ESPCN, we followed the default setting of neural networks in the reference(Shi et al., 2016a) and set upscale factor $\mathbf{r} = 3$. The take 70% of data samples for training and the remaining 30% samples for testing. The learning rate is set to 0.001. The training loss is converged after 2000 epochs. The ESPCN is implemented using PyTorch("PyTorch," n.d.), a popular scientific computing library. The training process takes roughly three hours on a NVIDIA TITAN X GPU using 3,800 training samples.

## 3. Results

**Fig. 2a** shows an example super-resolution result in comparison with the baseline bicubic upsampling applied to the raw SANS data. The input image is in the resolution of 30 by 30 pixels. High resolution prediction is 120 by 120, which is 16 times more pixels. The scattering image shown here exhibits one of the commonly observed scattering characteristics of SANS data. The scattering intensity is stongest at the lowest $q$ value and decays as the $q$ increses. Both bicubic algorithm and the proposed AI super-resolution method successfully predict the decaying intensity as a function of $q$ qualitatively well. However, noted that the AI predicted image not only produced the scattering intensity profile but also produced one of the instrument feature, a beamstop. In SANS experiments, beamstop is often used to block direct beam of neutrons. The beamstop absorbs neutrons, therefore it appears as empty pixels. At EQ-SANS, 60 mm diameter beamstop is used for 4m sample to detector distance configuration. The detector of EQ-SANS has an intrinsic pixel size of 5.5 mm x 4.3mm(Zhao et al., 2010). Therefore, the size of the beamstop corresponds to 10 - 15 pixels in width and height, respectively. However, the area of pixels shadowed by the beamstop is shown smaller in $q$-space due to the wide wavelength bands and the number of $q$-space bins. The pixel size of the low resolution image is estimated to be around ~20 mm in real space. Therefore, the beamstop does not show up in the low resolution $q$-space image, due to the scattering intensities contributed from different wavelength (**Fig. 2a. Low-resolution input**). Since there is no clear evidence of empty pixels representing a beamstop, bicubic algorithm fills the center pixels with predicted intensity. However, trained AI has identified the beamstop as one of the instrument feature and predicted positions of the empty pixels in the high resolution output (**Fig. 2a. AI**) even when the beamstop is not visually identified in the low resolution input. In **Fig. 2b**, the 1-D averaged scattering intensities from different sources are compared. Both results from bicubic algorithm and AI prediction exhibit reasonable agreement with the groundtruth. It is mainly due to the continuous and monotonic nature of this type of scattering profile, where bicubic algorithm can perform excellent. However, it should be pointed out that the AI predicted curve showed better accuracy especially at low-q ranges while the bicubic algorithm produced more deviations from the groundtruth than the AI model.

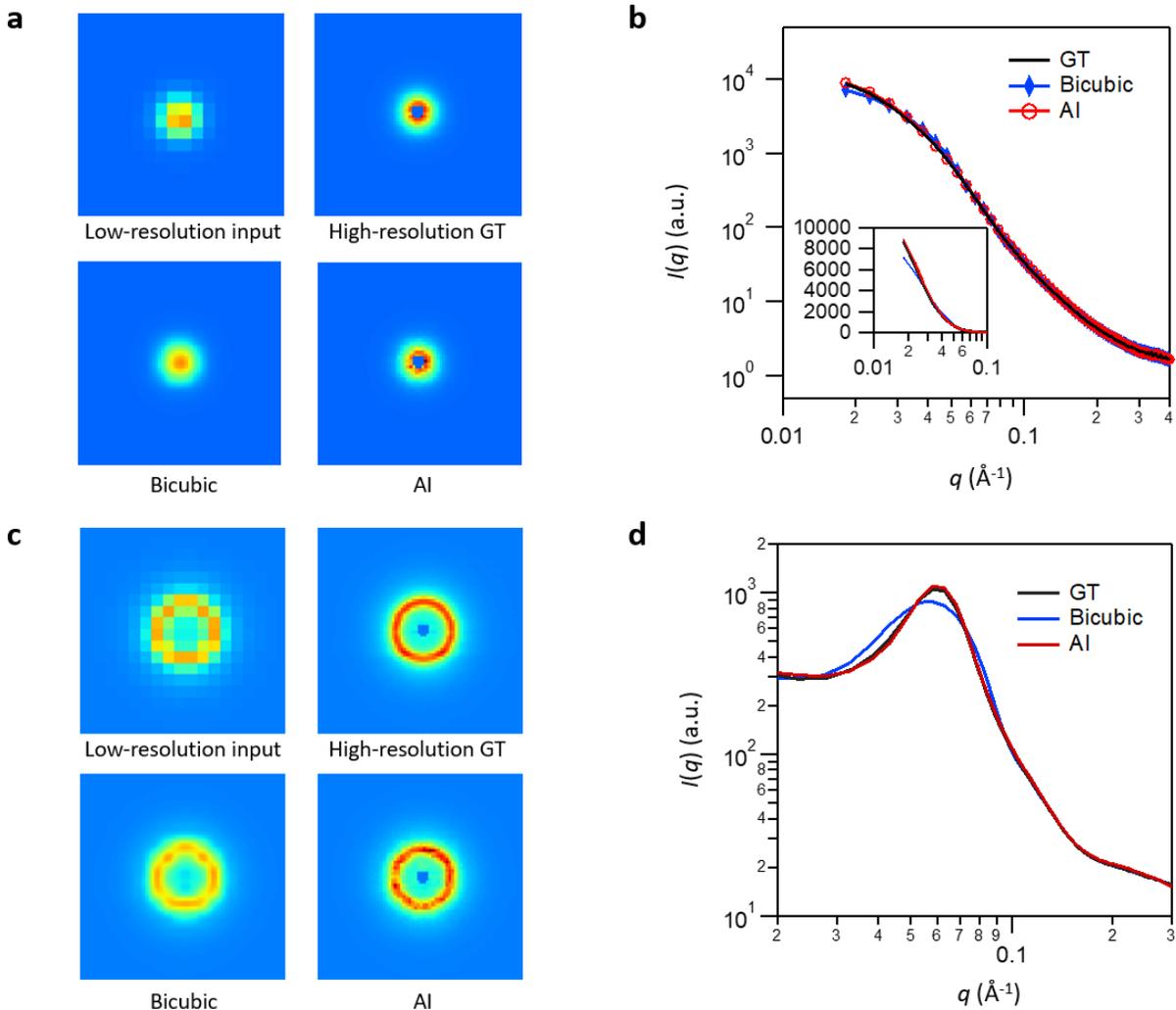

**Fig. 2.** Example of super-resolution AI on SANS data showing monotonic decay of scattering intensity as a function of $q$. (a) two-dimensional data showing low-resolution input, high-resolution ground truth (GT), high-resolution image predicted by bicubic algorithm, and high-resolution image predicted by AI model trained using EQ-SANS data. Images shown here are cropped around the centers of the full images to help visual comparison. (b) Circular averaged scattering profiles of the two-dimensional data shown in (a). Inset is the scaled-up plot of the low-q range with linear y-axis. (c) Example of two-dimensional SANS data with an isotropic peak. (d) Circular averaged scattering profiles of the data shown in (c).

We next apply the AI super-resolution model on another commonly encountered scattering pattern of SANS and compare the result with that from the bicubic algorithm. The isotropic ring pattern is often observed from SANS measurements, indicating strong correlation at certain length scale. For examples, interacting colloidal systems exhibit correlation peak whose width and sharpness can provide critical information about the distribution of particles and characteristics of interaction potentials. Randomly oriented liquid crystalline phase also produce scattering peaks which will appear as rings in two dimensional scattering pattern. The position and width of the rings can be used to extract structural dimensions and degree of order(Castro-Roman et al., 2005; Doe et al., 2009). Therefore, the super-resolution algorithm has to produce correct position and width of the such scattering features from the low-resolution input in order for the predicted scattering images to provide reliable information for further utilization. In **Fig. 2c**, results from the bicubic algorithm and the AI model are compared. Th AI super-resolution model has shown significantly better performance in recovering peak position and peak width compared to the bicubic algorithm. Bicubic algorithm resulted in distorted ring shape in 2D with much broader peak width, which would give incorrect physical interpretation. The difference of the peak shape can be clearly observed from the 1-D averaged scattering curve in **Fig. 2d**. The bicubic algorithm not only fails to predict correct width of the peak but also predicts the peak position slightly off. In contrast, AI model's prediction shows excellent agreement with the groundtruth data.

In SANS, anisotropic two-dimensional scattering patterns can be observed from samples that exhibit molecular level alignment and orientation. Soft materials under shear or stretching are well known examples that show such alignment, resulting in anisotropic scattering patterns(López-Barrón et al., 2017; Mortensen, 1996; Wang et al., 2017a). In principle, microstructures of deformed soft materials can be studied from the anisotropic scattering patterns, which can elucidate the relation between the structural deformation and resulting physical properties. Recent theoretical and experimental studies(Huang et al., 2017; Wang et al., 2017b) have also proven the importance of quantitative analysis on two dimensional scattering data for the fundamental understanding of rheological behavior. Our AI model has shown success in predicting high-resolution scattering pattern from the low-resolution anisotropic scattering data as well. **Fig. 3** shows typical anisotropic pattern of SANS measured at low-resolution and its high-resolution counterpart along with predictions by bicubic algorithm and AI model. From the 2D pattern, both high-resolution images reconstructed by the AI and bicubic algoritm show similar anisotropy found in the groundtruth image. This again shows that the bicubic algorithm works reasonably well when the scattering profile has slow and monotonic $q$ dependency without sharp transitions. The anisotropy represented by the integrated intensity from 0.03 Å$^{-1}$ < $q$ < 0.1 Å$^{-1}$ along the annulus also indicate the level of anisotropy is well-reproduced by both algorithms (**Fig. 4a**). One dimensional scattering intensities from the predicted 2D patterns are also compared by taking ±10º sectional average along the horizonal and vertical direction. In anisotropic scattering data analysis, scattering intensity profiles of specific orientation carry important information about the anisotropic structure of the sample. While both algorithms predicted the horizontal profiles with reasonably good agreement (**Fig. 4b**), AI prediction showed slightly better performance in reconstructing scattering intensity along the vertical direction (**Fig. 4c**).

In general, the AI model seems to be more accurate in predicting scattering images with sharp transitions. Since the bicubic algorithm relies on the smoothing and interpolation of data from one pixel to another, any lost information within a pixel cannot be recovered to a higher resolution image. On the other hand, the AI model is trained and equipped with a unique database that can connect the low-resolution pixel to pixel correlation information to a detailed high-resolution information. This enables for the AI model to grasp the attainable sharpness in the high-resolution training images, which may be described as an instrument resolution kernel.

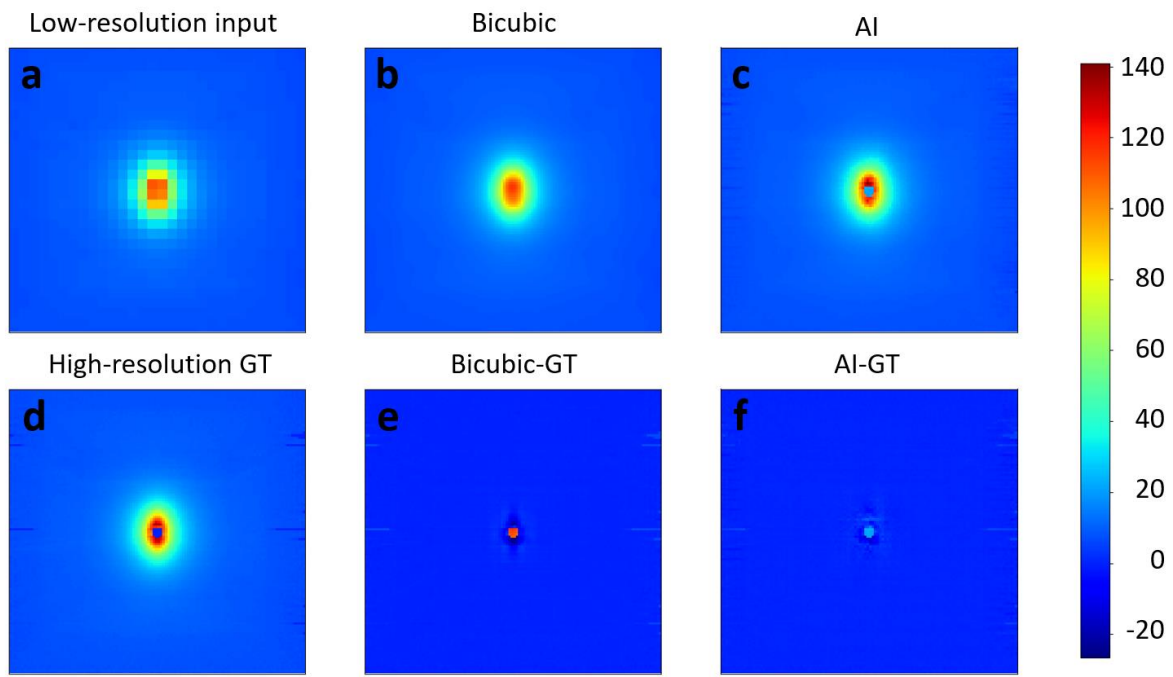

**Fig. 3**. Example of super-resolution process on a 2D anisotropic scattering image. (a) Anisotropic two-dimensional scattering image was used as an input to two-dimensional data showing low-resolution input for super-resolution algorithms. Results with enhanced resolution by (b) the bicubic algorithm and (c) AI model are shown. (d) High-resolution ground truth image. (e) and (f) Differences between the images obtained by the super-resolution algorithms and the ground truth.

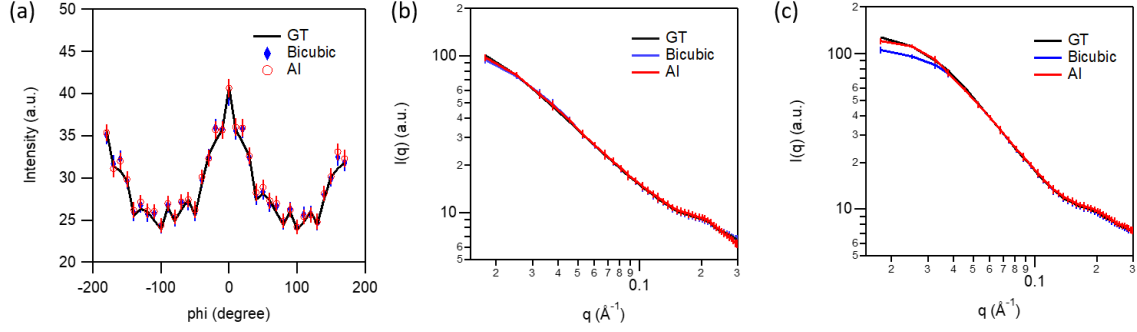

**Fig. 4**. (a) Annulus plot of integral intensities showing anisotropic character of the 2D data. (b) Sectional average along the horizonal direction of scattering patterns from **Fig. 3**. (c) Sectional average along the vertical direction of scattering patterns from **Fig. 3**.

In order to provide overall performance comparison between the AI model and the bicubic algorithm, differences between 1D averaged $I_{AI}(q)$ and $I_{GT}(q)$ have been estimated for all 1662 test cases using $\chi^2 = \Sigma\big(I_{AI\ or\ Bi}(q) - I_{GT}(q)\big)^2/N$, where $N$ is the number of $q$ values used in the summation. The $q$ range (0.02 Å$^{-1}$ < $q$ < 0.3 Å$^{-1}$) used in this calculation was chosen to exclude the shadowed pixels by the beamstop, because inclusion of these pixels will give huge penalty for the bicubic algorithm which do not produce the null pixels intrinsically. The results are summarized in **Fig. 5a** along with a dashed line indicating equal performance between the AI model and the bicubic model. It is quote clear that $\chi^2$ values from the bicubic model ($\chi^2_{Bi}$) are overall higher than the ones from the AI model ($\chi^2_{AI}$), suggesting that the AI model performs better than the bicubic algorithm. Closer look at the results is found in **Fig. 5b** where only the cases (1420 cases out of 1662) with at least one of the $\chi^2$ values being less than 10. A smaller value of the two $\chi^2$ values is denoted as $\chi^2_{min}$ and the other is denoted as $\chi^2_{max}$. This subset describes that quantitatively good prediction for the high-resolution scattering images has been achieved by one of the two algorithms. Red triangles indicate test cases where AI model's prediction was quantitatively better than the bicubic algorithm. And blue circle represents test cases where bicubic algorithm had smaller $\chi^2$. Interestingly, the distribution between red and blue data points can tell us about effectiveness of the AI model. The fact that blue data poitns are well concentrated in the $\chi^2_{max} < 10$ region shows that both AI and the bicubic models predicted the high-resolution images reasonably well and the different between the predictions was small. However, when AI was more successful than bicubic algorithm (red triangles in **Fig. 5b**), the $\chi^2_{max}$ values extends relatively large up to ~100. In other words, AI's prediction made much more significant improvements over the bicubic predictions.

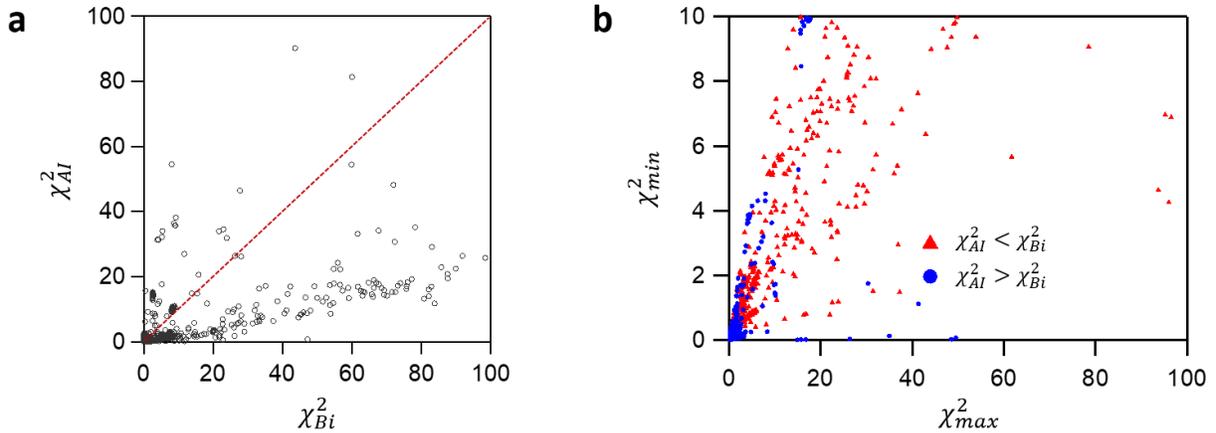

**Fig. 5**. (a) Distribution $\chi^2$ from the AI and bicubic super-resolution algorithm. Dashed line indicates the equal performance. (b) Distribution of $\chi^2$ values for cases with $\chi^2_{min} < 10$. Red triangles indicate cases where AI performed better than bicubic algorithm based on $\chi^2$ values. Blue circles represent cases where bicubic algorithm resulted in smaller $\chi^2$ values.

## 4. Discussion

We have introduced a novel way to speed up neutron scattering experiments by allowing data collection with low-spatial resolution and subsequently recovering the high-resolution data by using an AI model using DNN. By utilizing randomly selected SANS data from an identical instrument configuration, an AI super-resolution model has been trained. Our results demonstrate that the AI model trained in this way can successfully reconstruct high-resolution scattering images from the low-resolution data. Especially, the AI model has shown successful examples in restoring features such as scattering peaks with narrow width better than the traditional bicubic algorithm. The success may be attributed to the fact that the DNN model was able to capture instrument specific characteristics such as presence of the beamstop and instrument resolution which determines the broadening of scattering peaks during the training process. Being able to use low-resolution detector images is equivalent to reducing measurement time. In this study, we have shown that the high-resolution scattering data can be recovered from the low-resolution data with 16 times bigger pixels. Therefore, same level of counting statistsics can be achieved at 16 times faster rate. Considering that huge amount of cost and efforts are typically required to improve neutron flux or the performance of an instrument by 10 times, the approach proposed in this research may provide a new opportunities in neutron scattering sciences. If successfully integrated with existing beamlines after further studies on the AI models and training datasets, this method will not only accelerate the scientific process as it can help scientists' early decision making (*e.g.*, to stop the current scattering experiment if something is awry), but also provide new information for time-resolved scattering experiments at timescales which has not been possible before.

**Acknowledgements.** The Research at Oak Ridge National Laboratory's Spallation Neutron Source was sponsored by the Scientific User Facilities Division, Office of Basic Energy Sciences, U.S. Department of Energy.

**Corresponding Author** Correspondence and requests for materials should be addressed to C.D. (email: doc1@ornl.gov)